\documentstyle[preprint,tighten,eqsecnum,floats,aps,epsfig]{revtex}

\begin{document}

\preprint{nucl-th/9512026}
\draft

\title{ Relativistic Hamiltonians in many-body theories
}
\author{
 P. Amore, M.B. Barbaro, and A. De Pace
}
\address{
 $^a$ Dipartimento di Fisica Teorica dell'Universit\`a
 di Torino and \\
 Istituto Nazionale di Fisica Nucleare, Sezione di Torino, \\
 via P.Giuria 1, I-10125 Torino, Italy
}
\date{December 1995}

\maketitle

\begin{abstract}
We discuss the description of a many-body nuclear system using Hamiltonians
that contain the nucleon relativistic kinetic energy and potentials with
relativistic corrections. Through the Foldy-Wouthuysen transformation, the
field
theoretical problem of interacting nucleons and mesons is mapped to an
equivalent one in terms of relativistic potentials, which are then expanded at
some order in $1/m_N$. The formalism is applied to the Hartree problem in
nuclear matter, showing how the results of the relativistic mean field theory
can be recovered over a wide range of densities.
\end{abstract}

\vfill

\eject

\section{Introduction}
\label{sec:intro}

In recent years considerable efforts have been devoted to the development of
relativistic nuclear models, whose basis is usually a relativistic Lagrangian
containing nucleons and mesons\cite{Ser86,Cel86,Ser92}.
On the other hand, the traditional concept of interparticle potential has
proved
to be quite useful in nuclear physics and rather sophisticated techniques of
calculation have been developed within the hamiltonian formalism.

It is clear that it would be useful to be able to reformulate the relativistic
problem of interacting Dirac particles in terms of Pauli spinors and
potentials,
since this would allow one to employ the non-relativistic computational
techniques. Furthermore, the high energies and high momentum transfers
accessible at facilities such as Cebaf require calculations to be performed in
regimes where the relativistic kinematics is certainly important: then, one may
ask the question whether genuine relativistic dynamical effects are present and
to which extent they may reconducted to a potential treatment.

A possible way of connecting the two approaches relies on Hamiltonians
containing the nucleon relativistic kinetic energy and a potential
obtained by expanding at some order in $1/m_N$ the nucleon-nucleon (NN)
scattering amplitude.
For instance, in Ref.~\cite{Bar95a} the quasielastic charge response of
electron scattering has been evaluated in the relativistic Fermi gas model
using the Bonn potential expanded up to $1/m_N^2$. A similar approach has been
followed in Ref.~\cite{For95}, using a relativistic Hamiltonian in variational
Monte Carlo calculations.

The aim of the present paper is to show how a systematic expansion of the
nucleon-nucleon interaction in a many-body nuclear system can be constructed,
using the well-known Foldy-Wouthuysen transformation.
The many-body problem, expressed in terms of relativistic Green's functions and
meson-nucleon vertices, is transformed to an exactly equivalent one, in terms
of
modified Green's functions and vertices. After projection on the positive
energy
states and expansion in powers of $1/m_N$ one recovers the familiar
relativistic
potentials.

This procedure gives a systematic way of dealing with $1/m_N$ expansions in a
many-body context: it allows one to address, for instance, the issue of {\em
self consistency}, i.~e. the fact that the Dirac spinors depend on the nucleon
mass, which is modified by the medium.
In terms of relativistic Hamiltonians, this means that the potential, which
affects the effective nucleon mass via the self-energy, should in turn depend
on
it. In a fully relativistic context this problem is rather simple to solve,
whereas in a non-relativistic framework it is ill-defined.

In the following we shall not be concerned with antinucleons degrees of
freedom:
although this is an interesting issue in itself, we would like to maintain our
initial commitment of trying to understand the nuclear phenomena including
``simple'' relativistic effects, such as kinetic energies and
momentum-dependent
potentials. Any failure in this program might of course open the way to more
interesting relativistic effects, like, e.~g., vacuum polarization.

The reader should also be aware of a potential source of inconsistency
connected
to retardation effects. In fact, the parameters of empirical NN potentials,
such as the Bonn one, have usually been fixed assuming a static interaction.
Although the inclusion of retardation in the formalism developed in the present
paper is rather straightforward, when dealing with an empirical NN potential
this may not be the most appropriate course to follow.
Since in the following we have applied the formalism only to Hartree
calculations, which are not affected by this problem, we have considered, for
simplicity, static propagators for the meson fields.
The issue will have to be reconsidered when dealing, e.~g., with random phase
approximation calculations.

The paper is organized as follows: in Sect.~\ref{sec:rel-NN-pot}, the two most
commonly used schemes for defining a relativistic potential are briefly
sketched, employing $\sigma$ and $\omega$ exchange as examples.
In Sect.~\ref{sec:FW}, the Foldy-Wouthuysen expansion is introduced in the
nuclear many-body system, using nuclear matter for illustration, whereas in
Sect.~\ref{sec:rel-H-cal} the Hartree problem is solved, comparing the results
obtained with the relativistic Hamiltonians to the exact ones.
Finally, in the last Section we present our conclusions.

\section{Relativistic NN potentials}
\label{sec:rel-NN-pot}

In the literature one can find essentially two ways to derive a two-body
potential expanded up to a given order in $1/m_N$.

In the first case one expands the relativistic scattering amplitude and
interprets it as the matrix element of the Fourier transform of the potential
(Born term) between Pauli reduced spin wave functions.
We shall refer to it as ``Breit reduction''\cite{Lan71}.

An alternative procedure starts from the three-dimensional Blankenbecler-Sugar
(BbS) reduction of the Bethe-Salpeter (BS) equation, which yields an amplitude
satisfying elastic unitarity\cite{Bro69,Erk74}.

Both derivations are briefly sketched in the two following subsections. Here,
to
set the notation, let us define the momentum transfer $\bbox{q}$, the relative
momentum in the initial state $\bbox{p}$ and the total momentum $\bbox{P}$ for
the scattering from a two nucleon initial state with momenta $\bbox{k}_1$ and
$\bbox{k}_2$ to a final state with momenta $\bbox{k}'_1$ and $\bbox{k}'_2$:
\begin{mathletters}
\begin{eqnarray}
  \bbox{q} &=& \bbox{k}_1 - \bbox{k}'_1 = \bbox{k}'_2 - \bbox{k}_2 \\
  \bbox{p} &=& {\frac{1}{2}} \left( \bbox{k}_1 - \bbox{k}_2 \right) \\
  \bbox{P} &=& \bbox{k}_1 + \bbox{k}_2 = \bbox{k}'_1 + \bbox{k}'_2 \ .
\end{eqnarray}
\end{mathletters}
We will be considering the amplitudes for the exchange of a scalar particle of
mass $m_s$ (described by the scalar field $\phi$) and of a vector particle of
mass $m_v$ (described by the vector field $v_\mu$), whose couplings to the
nucleon are given by
\begin{mathletters}
\begin{eqnarray}
  H_s &=& g_s \overline{\Psi} \Psi \phi \\
  H_v &=& g_v \overline{\Psi} \gamma^\mu \Psi v_\mu \ ,
\end{eqnarray}
\end{mathletters}
where $\Psi$ is the nucleon field and $\gamma^\mu$ are the Dirac matrices.

\subsection{Breit reduction}
\label{subsec:Breit}

The amplitudes for the exchange of a scalar meson, $M_s$, and of a vector
meson, $M_v$, are given by:
\begin{mathletters}
\label{eq:rel-ampl}
\begin{eqnarray}
  M_s &=& - {\frac{g_s^2}{q^2 + m_s^2}} \overline{u} (\bbox{k}'_1)
    u (\bbox{k}_1) \overline{u} (\bbox{k}'_2) u (\bbox{k}_2) \\
  M_v &=& {\frac{g_v^2}{q^2 + m_v^2}} \overline{u} (\bbox{k}'_1)
    \gamma^\mu u (\bbox{k}_1) \overline{u} (\bbox{k}'_2) \gamma_\mu
    u (\bbox{k}_2) \ ,
\end{eqnarray}
\end{mathletters}
where $u (\bbox{k})$ is the Dirac spinor,
\begin{equation}
  u (\bbox{k}) = \sqrt{\frac{E_k + m_N}{2 m_N}}
    \left(
    \begin{array}{c}
      \chi \\
      {\frac{\bbox{\sigma} \cdot \bbox{k}}{E_k + m_N}} \chi
    \end{array}
    \right) \ ,
\end{equation}
whereas $\chi$ is the Pauli spinor and $E_k = \sqrt{k^2 + m_N^2}$.

When expanded up to order $k^2/m_N^2$, the normalized Dirac spinor turns out
to be\cite{Lan71}
\begin{equation}
  u (\bbox{k}) = \left( 1 - {\frac{k^2}{8 m_N^2}} \right)
    \left(
    \begin{array}{c}
      \chi \\
      {\frac{\bbox{\sigma} \cdot \bbox{k}}{2m_N}} \chi
    \end{array}
    \right)
    + O \left[ \left( {\frac{ k}{m_N}} \right)^4 \right]\ .
\end{equation}
Inserting this expression in (\ref{eq:rel-ampl}) and using the standard
relation between the Born amplitude and the spin matrix elements of the
potential in momentum space,
$M = {\chi'}_1^\dagger {\chi'}_2^\dagger V(\bbox{q},\bbox{p},\bbox{P})
 \chi_2 \chi_1$, one gets
\begin{mathletters}
\label{eq:VBR}
\begin{eqnarray}
  V^{\text{BR}}_s(\bbox{q},\bbox{p},\bbox{P}) &=&  - \frac{g_s^2}
    {q^2+m_s^2} \left\{ 1 - \frac{P^2}{4 m_N^2} -
    \frac{\left(2\bbox{p}-\bbox{q}\right)^2}{4 m_N^2}  \right. \nonumber  \\
    && \left. - i \bbox{p} \cdot \frac{\left[ \bbox{q} \times
    \left(\bbox{\sigma_1} + \bbox{\sigma_2} \right)\right]}{4 m_N^2} -
    i \bbox{P} \cdot \frac{\left[ \bbox{q} \times
    \left(\bbox{\sigma_1} - \bbox{\sigma_2} \right)\right]}{8 m_N^2} \right\}
\label{eq:VBR_sigma}    \\
  V^{\text{BR}}_v(\bbox{q},\bbox{p},\bbox{P}) &=& \frac{g_v^2}{q^2+m_v^2}
    \left\{ 1 - \frac{q^2}{4 m_N^2} -
    \frac{P^2}{4 m_N^2} + \frac{\left(2\bbox{p}-\bbox{q}\right)^2}{4 m_N^2}
    \right. \nonumber \\
    && \left. + 3 i \bbox{p} \cdot \frac{\left[ \bbox{q} \times
    \left(\bbox{\sigma}_1 + \bbox{\sigma}_2 \right)\right]}{4 m_N^2}
    - i \bbox{P} \cdot \frac{\left[ \bbox{q} \times
    \left(\bbox{\sigma}_1 - \bbox{\sigma}_2 \right)\right]}{8 m_N^2}  \right.
    \nonumber  \\
    && \left. - \frac{q^2}{6 m_N^2} \bbox{\sigma}_1 \cdot
    \bbox{\sigma}_2 + \frac{q^2}{12 m_N^2} S_{12} \right\}     \ ,
\end{eqnarray}
\end{mathletters}
where the superscript BR stands for ``Breit reduced'' and $S_{12}$ is the
standard tensor operator $S_{12} =  3 (\bbox{\sigma}_1\cdot\bbox{q})
(\bbox{\sigma}_2\cdot\bbox{q})/q^2 - \bbox{\sigma}_1 \cdot
\bbox{\sigma}_2$.

\subsection{Minimal relativity}
\label{subsec:min-rel}

A natural framework for relativistic two-nucleon potential scattering is given
by the BS equation. This is a four-dimensional integral equation for the
scattering amplitude, which is rather hard to solve even in the ladder
approximation. Hence, many approximation schemes have been derived in order to
obtain from it a more tractable three-dimensional, covariant equation, based on
the requirement of relativistic elastic unitarity\cite{Bro69,Erk74}.

A popular reduction procedure leads to the BbS equation, which, in the
centre-of-mass (c.m.) frame, reads
\begin{equation}
  M (\bbox{k}_f , \bbox{k}_i ) = V (\bbox{k}_f , \bbox{k}_i ) +
\int d \bbox{k} {\frac{m_N^2}{E_{k}}} V (\bbox{k}_f , \bbox{k} )
{\frac{ \Lambda_{+}^{(1)} (\bbox{k}) \Lambda_{+}^{(2)} (- \bbox{k})}
{k_i^2 - k^2 + i \epsilon}} M (\bbox{k} , \bbox{k}_i ) \ ,
\label{eq:BbS}
\end{equation}
where $V$ is the bare two-body potential, $\Lambda_{+}^{(i)}$ are the positive
energy projection operators and $\bbox{k}_i$, $\bbox{k}$ and $\bbox{k}_f$ are
the initial, intermediate and final c.m. momenta, respectively.

If one defines
\begin{mathletters}
\label{eq:min-rel}
\begin{eqnarray}
  \widetilde{M} (\bbox{k}_f,\bbox{k}_i ) &=& \sqrt{\frac{m_N}{E_{k_f}}}
    M (\bbox{k}_f , \bbox{k}_i ) \sqrt{\frac{m_N}{E_{k_i}}} \\
  \widetilde{V} (\bbox{k}_f,\bbox{k}_i ) &=& \sqrt{\frac{m_N}{E_{k_f}}}
    V (\bbox{k}_f , \bbox{k}_i ) \sqrt{\frac{m_N}{E_{k_i}}} \ ,
\end{eqnarray}
\end{mathletters}
Eq.~(\ref{eq:BbS}) becomes identical to the non-relativistic Lippmann-Schwinger
equation
\begin{equation}
  \widetilde{M} (\bbox{k}_f , \bbox{k}_i ) =
    \widetilde{V} (\bbox{k}_f , \bbox{k}_i ) +
    m_N \int d \bbox{k}  \widetilde{V} (\bbox{k}_f , \bbox{k} )
    {\frac{ \Lambda_{+}^{(1)} (\bbox{k}) \Lambda_{+}^{(2)} (- \bbox{k})}
    {k_i^2 - k^2 + i \epsilon}} \widetilde{M}
    (\bbox{k} , \bbox{k}_i ) \ .
\label{eq:Lippmann}
\end{equation}
Prescription (\ref{eq:min-rel}) is usually known as ``minimal
relativity''\cite{Bro69}.
It can be generalized to a generic frame by introducing a factor
$(m_N/E)^{\frac{1}{4}}$ for each spinor in the amplitude, i.e.
\begin{equation}
 V^{MR} (\bbox{q} , \bbox{p} ,\bbox{P} ) =
\left[ {\frac{m_N}{E_{k_1}}} {\frac{m_N}{E_{k'_1}}}
{\frac{m_N}{E_{k_2}}} {\frac{m_N}{E_{k'_2}}}
\right]^{\frac{1}{4}}  V (\bbox{q} , \bbox{p} ,\bbox{P} ) \ ,
\end{equation}
where MR stands for ``minimal relativity''.

Expanding again up to the order $k^2/m_N^2$ one gets, for the exchange of
scalar
and vector mesons, the following expressions
\begin{mathletters}
\label{eq:VMR}
\begin{eqnarray}
  V_s^{\text{MR}}(\bbox{q},\bbox{p},\bbox{P}) &=&
    - \frac{g_s^2}{q^2+m_s^2} \left\{ 1 -
    \frac{P^2}{8 m_N^2} - \frac{\left(2\bbox{p}-\bbox{q}\right)^2}{8 m_N^2}
    + \frac{q^2}{8 m_N^2} + \right. \nonumber \\
    && \left. - i \bbox{p} \cdot \frac{\left[ \bbox{q} \times
    \left(\bbox{\sigma}_1 + \bbox{\sigma}_2 \right) \right]}{4 m_N^2}
    - i \bbox{P} \cdot \frac{\left[ \bbox{q} \times
    \left(\bbox{\sigma}_1 - \bbox{\sigma}_2 \right) \right]}{8 m_N^2}
    \right\}      \\
  V_v^{\text{MR}}(\bbox{q},\bbox{p},\bbox{P}) &=&
    \frac{g_v^2}{q^2+m_v^2}
    \left\{ 1 - \frac{q^2}{8 m_N^2} - \frac{P^2}{8 m_N^2} + 3
    \frac{\left(2\bbox{p}-\bbox{q}\right)^2}{8 m_N^2} + \right. \nonumber \\
    && \left. + 3 i \bbox{p} \cdot \frac{\left[ \bbox{q} \times
    \left(\bbox{\sigma}_1 + \bbox{\sigma}_2 \right) \right]}{4 m_N^2}
    - i \bbox{P} \cdot \frac{\left[ \bbox{q} \times
    \left(\bbox{\sigma}_1 - \bbox{\sigma}_2 \right) \right]}{8 m_N^2}
    - \right. \nonumber \\
    && \left. - \frac{q^2}{6 m_N^2} \bbox{\sigma}_1 \cdot
    \bbox{\sigma}_2 + \frac{q^2}{12 m_N^2} S_{12}
    \right\}     \ .
\end{eqnarray}
\end{mathletters}
Eq.~(\ref{eq:Lippmann}) has been used in deriving the relativistic Bonn
potential\cite{Mac87}. Hence, the non-relativistic expansion of this potential,
useful in nuclear structure calculations, is given, for $\sigma$ and $\omega$
exchange, by Eq.~(\ref{eq:VMR}).

\section{Foldy-Wouthuysen expansion}
\label{sec:FW}

In this section we would like to rederive the previous expansions using the
language of the Green's functions and within the scheme devised by Foldy and
Wouthuysen\cite{Fol50} (FW) to decouple the large and small components in the
relativistic wave function.

Let us start by introducing the FW unitary transformation\cite{Bjo64,Itz80},
\begin{equation}
    {\cal T} (\bbox{k}) = \sqrt{\frac{E_{k} + m_N}{2 E_{k}}}
     \left( \openone + \frac{\bbox{\gamma} \cdot \bbox{k}}{E_{k}
     + m_N} \right)  \ ,
\label{eq:FWtr}
\end{equation}
and the associated FW nucleonic field operator,
\begin{equation}
  \begin{array}{rcl}
    \Psi^{\text{FW}}             &=&  {\cal T} \Psi            \\
    \overline{\Psi}^{\text{FW}}  &=&  \overline{\Psi} {\cal T}
  \end{array}
\label{eq:FWave}
\end{equation}
(having used the property $\gamma^0 {\cal T}^\dagger \gamma^0 = {\cal T}$).

Using (\ref{eq:FWave}) one can define a FW Green's function
\begin{eqnarray}
\label{eq:FWgr}
  i G^{\text{FW}} (x,y) &\equiv&
    \langle\Psi_0|\text{T}\left[\Psi^{\text{FW}}(x)
  \overline{\Psi}^{\text{FW}}(y) \right] |\Psi_0\rangle  \nonumber \\
                      &=& {\cal T} i G ( x , y ) {\cal T} \ .
\end{eqnarray}
In momentum space, and for the case of the relativistic Fermi gas, the latter
reads
\begin{eqnarray}
   G_0^{\text{FW}} (k,k_0) &\equiv& {\cal T}(\bbox{k}) G_0 (k,k_0)
     {\cal T}(\bbox{k})  \nonumber \\
                     &=& P_{+} \left[ {\frac{\vartheta (k-k_F)}{k_0 -
 E_{k} + i \varepsilon}} + {\frac{\vartheta (k_F - k)}{k_0 -
 E_{k} - i \varepsilon}} \right] - {P_{-}} {\frac{1}{k_0 +
 E_{k} - i \varepsilon}} \ ,
\end{eqnarray}
where $P_{\pm}\equiv(\openone\pm\gamma^0)/2$ are operators that project out the
large/small components in the wave function and $k_F$ is the Fermi momentum.

It is evident that one gets two pieces, acting separately on the large and
small
components, since the transformation (\ref{eq:FWtr}) has washed out all the
operators inducing a mixing (i.e. the $\gamma_i$'s).

\begin{figure}[tb]
\begin{center}
\mbox{\epsfig{file=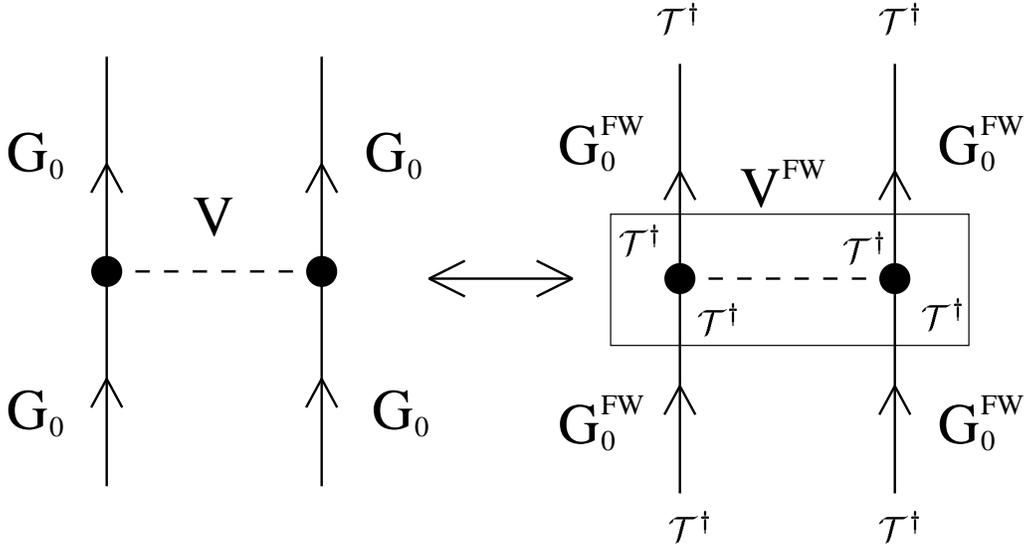}}
\vskip 2mm
\caption{ Scattering amplitude in terms of bare and FW propagators and
potentials.
  }
\label{fig:FW-interaction}
\end{center}
\end{figure}

It is also clear from the definition (\ref{eq:FWgr}) that one can always
redefine any given Feynman diagram in terms of $G^{\text{FW}}$ and a
transformed
interaction (see Fig.~\ref{fig:FW-interaction}).

For a typical meson-exchange interaction one has, for instance,
\begin{equation}
V = g^2 \Gamma_1 D (q) \Gamma_2 \ ,
\end{equation}
where $g$ is the coupling constant, $D (q)$ the meson propagator and $\Gamma_i$
the vertex operators (e.g., $\openone$ or $\gamma^\mu$).

Then, the FW interaction would read
\begin{equation}
  V^{\text{FW}} = g^2 {\cal T}^\dagger(\bbox{k}_1')
    \Gamma_1 {\cal T}^\dagger(\bbox{k}_1) D(q) {\cal T}^\dagger(\bbox{k}_2')
    \Gamma_2 {\cal T}^\dagger(\bbox{k}_2) \ .
\label{eq:pot}
\end{equation}
We make now a further simplification, neglecting in the following the
contribution of antinucleons:
\begin{equation}
  G_0^{\text{FW}}(k,k_0) \rightarrow G_{0 (+)}^{\text{FW}}(k,k_0) =
  P_{+} \left[ {\frac{\vartheta (k-k_F)}{k_0 -
  E_{k} + i \varepsilon}} + {\frac{\vartheta (k_F - k)}{k_0 -
  E_{k} - i \varepsilon}} \right]  \ .
\end{equation}
It is a simple matter to move the operators $P_{+}$ from the Green's functions
to the potential, defining
\begin{equation}
  V_{(+)}^{\text{FW}} = g^2 P_{+}
  {\cal T}^\dagger(\bbox{k}_1') \Gamma_1
  {\cal T}^\dagger(\bbox{k}_1) P_{+} D(q) P_{+} {\cal T}^\dagger(\bbox{k}_2')
  \Gamma_2 {
\cal T}^\dagger(\bbox{k}_2) P_{+}  \ .
\label{eq:pot+}
\end{equation}

The many-body problem is then formally equivalent to the standard
non-relativistic one, in terms of a potential $V_{(+)}^{\text{FW}}$ and the
Green's function
\begin{equation}
  G_{0(+)}(k,k_0) = \left[ {\frac{\vartheta (k-k_F)}{k_0 -
    E_{k} + i \varepsilon}} + {\frac{\vartheta (k_F - k)}{k_0 -
    E_{k} - i \varepsilon}} \right]  \ ,
\label{eq:G0+}
\end{equation}
which is identical to the non-relativistic one, apart from the energy-momentum
relation.

For the case of scalar and vector meson exchange, the FW reduced potentials
have
the following form:
\begin{mathletters}
\label{eq:VFW}
\begin{eqnarray}
  V_{s (+)}^{\text{FW}} &=&
    -\frac{g_s^2}{q^2+m_s^2} \sqrt{\frac{(E_{k'_1}+m_N)
    (E_{k_1}+m_N) (E_{k_2'}+m_N) (E_{k_2}+m_N)}
    {16 E_{k'_1} E_{k_1} E_{k_2'} E_{k_2}}}
    \nonumber \\
  && \times \left\{ 1 -
    \frac{\bbox{k}_1 \cdot \bbox{k}'_1}{(E_{k'_1}+m_N)(E_{k_1}+m_N)}
    - \frac{i \bbox{k}'_1 \cdot \left( \bbox{k}_1 \times
    \bbox{\sigma}_1 \right)}{(E_{k'_1}+m_N)(E_{k_1}+m_N)}  \right\}
    \nonumber \\
  && \times \left\{ 1 -
    \frac{\bbox{k}_2 \cdot \bbox{k}'_2}{(E_{k_2'}+m_N)(E_{k_2}+m_N)}
    - \frac{i \bbox{k}'_2 \cdot \left( \bbox{k}_2 \times
    \bbox{\sigma}_2 \right)}{(E_{k_2'}+m_N)(E_{k_2}+m_N)}  \right\}
\label{eq:VFW_sigma} \\
  V_{v (+)}^{\text{FW}} &=&
    \frac{g_v^2}{q^2+m_v^2}
    \sqrt{\frac{(E_{k'_1}+m_N) (E_{k_1}+m_N)(E_{k_2'}+m_N) (E_{k_2}+m_N)}
    {16 E_{k'_1} E_{k_1} E_{k_2'} E_{k_2}}}
    \nonumber \\
  && \times \left\{ \left[ 1 +
    \frac{\bbox{k}_1 \cdot \bbox{k}'_1}{(E_{k'_1}+m_N)(E_{k_1}+m_N)}
    + \frac{i \bbox{k}'_1 \cdot \left( \bbox{k}_1 \times
    \bbox{\sigma}_1 \right)}{(E_{k'_1}+m_N)(E_{k_1}+m_N)}  \right] \right.
    \nonumber \\
  && \left. \quad \times \left[ 1 +
    \frac{\bbox{k}_2 \cdot \bbox{k}'_2}{(E_{k_2'}+m_N)(E_{k_2}+m_N)}
    + \frac{i \bbox{k}'_2 \cdot \left( \bbox{k}_2 \times
    \bbox{\sigma}_2 \right)}{(E_{k_2'}+m_N)(E_{k_2}+m_N)}  \right] \right.
    \nonumber \\
  && \left. \quad + \left[ i \left( \frac{\bbox{k}'_1}{E_{k'_1}+m_N} -
    \frac{\bbox{k}_1}{E_{k_1}+m_N} \right) \times \bbox{\sigma}_1 -
    \left( \frac{\bbox{k}_1}{E_{k_1}+m_N} +
    \frac{\bbox{k}'_1}{E_{k'_1}+m_N} \right) \right] \right.
    \nonumber \\
  && \left. \quad\ \cdot \left[ i \left(\frac{\bbox{k}_2}{E_{k_2}+m_N} -
    \frac{\bbox{k}'_2}{E_{k_2'}+m_N} \right) \times \bbox{\sigma}_2 +
    \left( \frac{\bbox{k}_2}{E_{k_2}+m_N} +
    \frac{\bbox{k}'_2}{E_{k_2'}+m_N} \right) \right] \right\} \ .
\end{eqnarray}
\end{mathletters}
It is then a straightforward matter to check, by expanding these expressions up
to second order in the inverse nucleon mass, that one gets the $V^{\text{BR}}$
potentials of Eq.~(\ref{eq:VBR}).
On the other hand, if one performs the expansions dropping in (\ref{eq:VFW})
the big square root containing kinematical factors $(E_k+m_N)/2E_k$, then one
recovers the $V^{\text{MR}}$ potentials of Eq.~(\ref{eq:VMR}).

The kinematical factors of the minimal relativity prescription
(\ref{eq:min-rel}) have been introduced in order to reduce the BS
equation to a non-relativistic Lippmann-Schwinger equation: hence, it would
appear natural to use the non-relativistic expression for the kinetic energy in
doing calculations with the $V^{\text{MR}}$ potentials. Note, however, that in
the literature both the non-relativistic\cite{Bro69,Haj80} and the
relativistic\cite{Hol81,Mac86} kinetic energies have been employed.

\section{Relativistic Hartree calculations}
\label{sec:rel-H-cal}

We have seen that the FW transformation leads to a new Green's function,
in which the small and the large components are decoupled.
Simultaneously, we define the potentials according to the prescription
(\ref{eq:pot+}). In this way we are able to operate within a framework that
clearly resembles the non-relativistic one: the FW Green's function,
projected upon the large components, has indeed the same structure of the
familiar non-relativistic Green's function (see (\ref{eq:G0+})), apart from the
energy of the fermion, which is now fully relativistic.
On the other hand, the potentials obtained from a non-relativistic expansion
of $V^{\text{FW}}$ reproduce the ones obtained following other procedures
(for example, see \cite{For95,Mac87,Haj80}).

Let us now go further and apply the FW transformation to the relativistic
Hartree approximation for the nucleon Green's function, assuming again that
only scalar and vector isoscalar mesons are exchanged among the nucleons.
This is the basis of many relativistic nuclear structure calculations.
We will compare the results in the present approach with those obtained in the
mean field theory\cite{Ser86} (MFT), i.e. the Hartree approximation in which
antinucleon contributions to the self-energies are neglected.

In the MFT the vector field shifts the energy of the nucleon, while the
scalar field dresses the mass. It is worth stressing that this property follows
directly from the Lorentz structure of the self-energy: the scalar self-energy,
which is a scalar under Lorentz transformations, transforms as the mass, while
the vector self-energy transforms as the time component of a four-vector.
An important related issue in a Hartree calculation is connected to
{\em self-consistency}. Indeed, in a relativistic framework the spinors depend
on the nucleon mass, which is modified in the medium and should then be
calculated self-consistently. This is rather straightforward in the MFT,
because
of the above mentioned Lorentz structure of the self-energy. In a
non-relativistic calculation, based on an effective potential expanded at some
order in $k^2/m^2$, the dependence on the nucleon mass has been shifted to the
potential: however, in this case one can no longer rely on general symmetry
arguments and the FW expansion provides a systematic way to deal with
self-consistency.

Let us start by writing the Dyson equation for the inverse Hartree Green's
function
\begin{equation}
G_{\text{H}}(k,k_0)^{-1} = G_{0}(k,k_0)^{-1} - \Sigma_{\text{H}} \ ,
\label{eq:Dys-Har}
\end{equation}
where $G_{0}(k,k_0)^{-1} = \gamma^0 k_0-\bbox{\gamma}\cdot\bbox{k} - m_N$ is
the inverse of the free Green's function, whereas $\Sigma_{\text{H}}$ is the
Hartree self-energy due to the exchange of $\sigma$ and $\omega$, i.~e.
\begin{eqnarray}
  \Sigma_{\text{H}} &\equiv& \Sigma_{\text{H}}^{s} - \gamma_0
    \Sigma_{\text{H}}^{v}  \nonumber \\
  &=& i \frac{g_s^2}{m_s^2} \int \frac{d^4 k}{(2 \pi)^4} \text{Tr}
    \left[ G_{\text{H}} (k,k_0) \right] e^{i k_0 \eta}
    - i \frac{g_v^2}{m_v^2} \int \frac{d^4 k}{(2 \pi)^4}\gamma_0\text{Tr}
    \left[ \gamma^0 G_{\text{H}} (k,k_0) \right] e^{i k_0 \eta} \ .
\label{eq:self-energies}
\end{eqnarray}
The solution of (\ref{eq:Dys-Har}) is clearly:
\begin{equation}
  G_{\text{H}}(k,k_0)^{-1} = \gamma^0\overline{k}_0-\bbox{\gamma}\cdot\bbox{k}
    - m_N^{*},
\end{equation}
with
\begin{mathletters}
\label{eq:self-cons-MFT}
\begin{eqnarray}
    \overline{k}_0 &=& k_0 + \Sigma_{H}^{v} \\
    m_N^* &=& m_N+\Sigma_{\text{H}}^{s}
\label{eq:mnstar}
\end{eqnarray}
\end{mathletters}
being the nucleon energy shifted by the vector field and the nucleon mass
dressed by the scalar field, respectively. These relations, together with
Eq.~(\ref{eq:self-energies}), define the self-consistent Hartree solution in
the MFT.

Applying the FW procedure, one can introduce in the same way a Dyson equation
for the FW Hartree Green's function:
\begin{equation}
  G_{\text{H}}^{\text{FW}} (k,k_0)^{-1} =
    {\cal T}_{\text{H}}^\dagger(\bbox{k})
    {\cal T} (\bbox{k}) G_{0}^{\text{FW}} (k,k_0)^{-1} {\cal T} (\bbox{k})
    {\cal T}_{\text{H}}^\dagger (\bbox{k}) -
    {\cal T}_{\text{H}}^\dagger (\bbox{k}) \Sigma_{\text{H}}
    {\cal T}_{\text{H}}^\dagger (\bbox{k}) \ ,
\label{eq:Dys-FW}
\end{equation}
where $G_{0}^{\text{FW}}(k,k_0)^{-1} = \gamma^0 k_0 - E_k$ is the inverse of
the free FW Green's function, while ${\cal T}_{\text{H}}(\bbox{k})$ is the
operator (\ref{eq:FWtr}), evaluated for a nucleon with dressed mass $m_N^*$.
The solution to this equation can again be written as
\begin{equation}
  G_{\text{H}}^{\text{FW}}(k,k_0)^{-1} = \gamma^0\overline{k}_0 - E_k^{*},
\end{equation}
with $E_k^{*}=\sqrt{k^2+{m_N^*}^2}$ and $\overline{k}_0$ and $m_N^*$ still
given by Eq.~(\ref{eq:self-cons-MFT});
that is, at this stage one recovers the results of MFT.

Now, in order to make contact with the effective potentials of the previous
Sections, one has to project away the small components and to express the
self-energies in terms of the potentials.
Starting again from Eq.~(\ref{eq:Dys-FW}), the left hand side becomes
\begin{equation}
  P_+ G_{\text{H}}^{\text{FW}}(k,k_0)^{-1} P_+ = P_+ (\overline{k}_0 - E_k^{*})
    \equiv P_+ G_{\text{H}(+)}(k,k_0)^{-1} \ ,
\end{equation}
whereas the last term on the right hand side can be rewritten as
\begin{eqnarray}
  P_+ {\cal T}_{\text{H}}^\dagger (\bbox{k})
   \Sigma_{\text{H}} {\cal T}_{\text{H}}^\dagger (\bbox{k}) P_{+}
   &=& - i \int \frac{d^4 p}{(2 \pi)^4} P_+ \text{Tr} \left[
   V^{\text{FW}}_{H(+)} (\bbox{k},\bbox{p}) G_{H(+)}^{\text{FW}} (p,p_0)\right]
   e^{i p_0 \eta} \nonumber \\
  &=& P_+\left(\frac{m_N^{*}}{E_k^{*}} \Sigma_{\text{H}}^{s} -
    \Sigma_{H}^{v}\right) \nonumber \\
  &\equiv& P_+ \Sigma^{\text{FW}}_{H(+)} \ .
\end{eqnarray}
In the previous equation we have introduced for brevity $V^{\text{FW}}_{H(+)}
(\bbox{k},\bbox{p})\equiv V^{\text{FW}}_{(+)}(\bbox{q}=0,(\bbox{k}-\bbox{p})/2,
\bbox{k}+\bbox{p})$.
Then, Eq.~(\ref{eq:Dys-FW}) becomes
\begin{equation}
  \overline{k}_0 - E_{k}^{*} = \left( k_0 + \Sigma_{H}^{v} \right)
    - \frac{k^2 + m_N^{*}(m_N + \Sigma_{\text{H}}^{s})}{E_k^*} \ ,
\label{eq:Dys2}
\end{equation}
where we have gathered separately the terms shifting the energy and the terms
dressing the mass.
The self-consistency condition is still given by Eq.~(\ref{eq:mnstar}), as one
can easily check. However, we can now reformulate it in terms of the FW
potentials, namely
\begin{equation}
  m_N^{*} = m_N - i \frac{E_k^{*}}{m_N^{*}}
    \int \frac{d^4 p}{(2 \pi)^4} \text{Tr} \left[
    V^{\text{FW}}_{H(+)}(\bbox{k},\bbox{p})^{s}
    G_{H(+)}^{\text{FW}} (p,p_0) \right] e^{i p_0 \eta} \ .
\label{eq:self1}
\end{equation}
If we expand the potential up to $O(1/m_N^4)$, then we
obtain an approximate self-consistency equation for the effective mass:
using the $\sigma$ potential of Eq.~(\ref{eq:VFW_sigma}), that is, after
expansion, the $V^{\text{BR}}_s$ potential of Eq.~(\ref{eq:VBR_sigma}), we get
\begin{equation}
  m_N^{*} = m_N - \rho \frac{g_s^2}{m_s^2} \left( 1 - \frac{3 k_F^2}{10
    {m_N^{*}}^2} \right) \ ,
\label{eq:self2}
\end{equation}
where $\rho = 2 k_F^3/3 \pi^2$ is the density of nuclear matter.

\begin{figure}[tb]
\begin{center}
\mbox{\epsfig{file=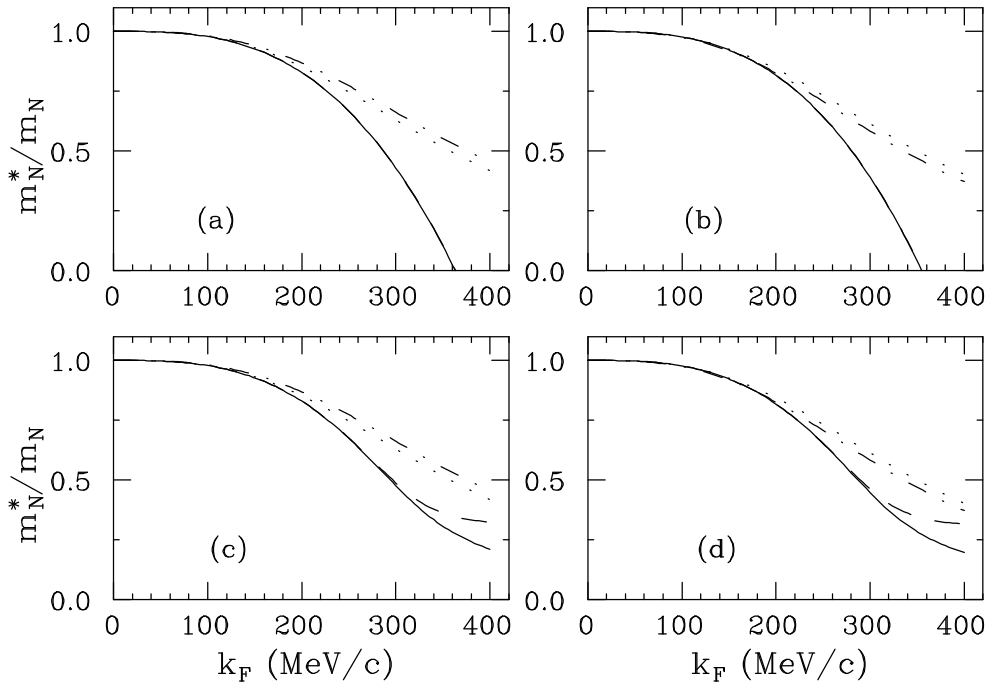}}
\vskip 2mm
\caption{ The effective mass as a function of the Fermi momentum for the MFT of
Eq.~(\protect\ref{eq:mnstar}) (solid), the approximation
(\protect\ref{eq:self2}) (dashed) and the non-relativistic models
(\protect\ref{eq:mnBR}) and (\protect\ref{eq:mnMR}) (dotted and dot-dashed,
respectively). The upper panels correspond to the calculation neglecting
self-consistency (see text), whereas in the lower ones the relativistic
$m_N^*$'s are self-consistent. The left hand panels correspond to the MFT
choice of parameters and the right hand ones to the Bonn potential parameters.
Note that in (a) and (b) the solid and dashed lines practically coincide.
  }
\label{fig:mnstar-rel}
\end{center}
\end{figure}

The exact expression (\ref{eq:mnstar}) and the approximate one (\ref{eq:self2})
may be compared with the standard non-relativistic formula for $m_N^*$:
\begin{equation}
  m_N^* = \frac{m_N}{\displaystyle 1+\frac{\strut m_N}{\displaystyle k}
    \frac{\strut\partial\Sigma}{\displaystyle\partial k}}.
\end{equation}
Note that from this formula one gets a constant effective mass in the Hartree
approximation only when the interaction is quadratic in the non-local part,
whereas (\ref{eq:self2}) yields a constant $m_N^*$ at any order.
Using the $V^{\text{BR}}$ and the $V^{\text{MR}}$ potentials of
Sect.~\ref{sec:rel-NN-pot} one gets
\begin{mathletters}
\label{eq:mnstar-nonrel}
\begin{eqnarray}
  m_N^* &=& \frac{m_N}{\displaystyle 1+ \frac{\strut\rho}{\displaystyle m_N}
    \frac{\strut g_s^2}{\displaystyle m_s^2}} \qquad\qquad\qquad\quad
    \text{Breit reduced}
\label{eq:mnBR} \\
  m_N^* &=& \frac{m_N}{\displaystyle 1+ \frac{\strut\rho}{\displaystyle 2m_N}
    \left( \frac{\strut g_s^2}{\displaystyle m_s^2} +
           \frac{\strut g_v^2}{\displaystyle m_v^2} \right)}
    \qquad \text{Minimal relativity} \ .
\label{eq:mnMR}
\end{eqnarray}
\end{mathletters}
It is apparent that while the $V^{\text{BR}}$ potential complies with the
relativistic symmetry requirements (i.~e., only $\sigma$ exchange dresses the
mass), in the case of $V^{\text{MR}}$ both $\sigma$ and $\omega$ exchange are
contributing.
We display in Fig.~\ref{fig:mnstar-rel} $m_N^*$ as a function of $k_F$ for
two choices of the coupling constants: one set is taken from the MFT
\cite{Ser86} (panels (a) and (c)), where they are fixed in order to reproduce
the saturation point of nuclear matter, and the other from the Bonn potential
\cite{Mac87} (panels (b) and (d)), which is fitted to the NN scattering
data\footnote{Note that the Bonn potential contains also form factors,
depending
on the momentum transfer, which could be easily accomodated in the formalism.
However, since in the Hartree self-energy ($q=0$) they produce only a slight
rescaling of the coupling constants, we shall neglect them in the following
discussion, which has only illustrative purposes. }.
The values of the parameters can be found in Table~\ref{tab:coupl-const}.
The largest difference is, of course, in the vector meson coupling, which is
artificially reduced in the MFT in order to simulate the effect of short-range
correlations.

\begin{table}
\caption{ Masses and coupling constants of the $\sigma$ and $\omega$ mesons
in the MFT \protect\cite{Ser86} and in the Bonn potential (BP)
\protect\cite{Mac87}.
 }
\label{tab:coupl-const}
\begin{tabular}{lcdcd}
      & $m_s$ (MeV) & $g_s^2/4\pi$ & $m_v$ (MeV) & $g_v^2/4\pi$ \\
  \tableline
  MFT & 550         & 7.29         & 783         & 10.84        \\
  BP  & 550         & 7.7823       & 783         & 20.           \\
\end{tabular}
\end{table}

In the case of the two upper panels ((a) and (b)), Eqs.~(\ref{eq:mnstar}) and
({\ref{eq:self2}}) have been solved dropping the requirement of
self-consistency, i.~e. utilizing the bare mass $m_N$ instead of $m_N^*$ in
the right hand side of these equations. In the lower panels ((c) and (d)) the
fully self-consistent solutions have been given. In the non-relativistic cases
we have always applied Eq.~(\ref{eq:mnstar-nonrel}).

First of all, one should notice the remarkably good agreement between the exact
expression (\ref{eq:mnstar}) for $m_N^*$ and the approximate one
({\ref{eq:self2}}) (in particular, in the non-self-consistent instance the two
curves are practically indistinguishable).
The non-relativistic approximation fails badly at the standard nuclear matter
density and above. Note, however, that when the Fermi gas is applied to
relatively light nuclei an effective lower value for $k_F$ should be employed.
For instance, in Ref.~\cite{Bar95a} an analysis of the charge response has been
performed, using the expanded Bonn potential, for $^{12}$C at $k_F=225$ MeV/c.
In that case, the non-relativistic $m_N^*$ differs from the exact one by less
than 3\% (Fig.~\ref{fig:mnstar-rel}d).

The dependence on the parameters is rather mild: while this is expected for the
relativistic calculations and for $V^{\text{BR}}$, which depend only on the
rather stable $\sigma$ coupling constant (see Table~\ref{tab:coupl-const}), it
turns out that also the dependence on $g_v^2$ in $V^{\text{MR}}$ is not
dramatic.

\begin{figure}[tb]
\begin{center}
\mbox{\epsfig{file=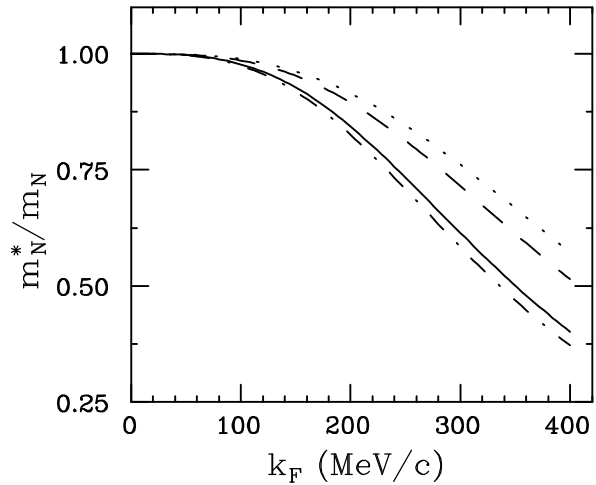}}
\vskip 2mm
\caption{ The effective mass as a function of the Fermi momentum for the
non-relativistic models (\protect\ref{eq:mnBR}) (solid) and
(\protect\ref{eq:mnMR}) (dot-dashed) using the Bonn potential set of
parameters. The effective mass stemming from $V^{\text{MR}}$ when only the
$\sigma$ (dotted) or the $\omega$ (dashed) mesons are employed is also shown.
  }
\label{fig:mnstar-nonrel}
\end{center}
\end{figure}

It is worth noticing that in spite of the fact that the effective mass
stemming from $V^{\text{MR}}$ violates the relativistic symmetry requirements,
the numerical values turn out to be rather close to the ones obtained from
$V^{\text{BR}}$: the contribution of the $\sigma$ meson is reduced to make room
for the $\omega$ meson. We compare in Fig.~\ref{fig:mnstar-nonrel} the
effective
masses (\ref{eq:mnstar-nonrel}), showing also the individual contribution of
$\sigma$ and $\omega$ in the case of $V^{\text{MR}}$.

Coming back to the approximate relativistic expression ({\ref{eq:self2}}), it
is
clear from Figs.~\ref{fig:mnstar-rel}c and \ref{fig:mnstar-rel}d that the
expansion breaks above $k_F\approx300$ MeV/c. While this is enough for most
nuclear matter calculations, it is worth pointing out that the range of
validity
can be increased by considering higher order contributions.
Note that for momentum space calculations, such as those of Ref.~\cite{Bar95a},
the use of a potential expanded up to an arbitrary order does not pose any
major problem.

\section{Conclusions}
\label{sec:concl}

In this paper we have presented a systematic way of dealing with $1/m_N$
expansions in a relativistic many-body nuclear system.
The formalism has been applied to the Hartree problem, showing the importance
of
accounting, at the same order in the expansion, also for the condition of
self-consistency. Treating the expanded potentials strictly
non-relativistically
can lead to substantial discrepancies with the relativistic calculation,
depending on the density. On the other hand, a correct treatment of the
self-consistency condition considerably enlarges the range of densities where
the relativistic potentials are applicable, already at order $1/m_N^2$.

A word of caution should be spent about the use of an empirical NN interaction,
such as the Bonn potential: as we have seen, the minimal relativity
prescription, which is used in deriving the potential, does not appear to be
consistent with a relativistic dispersion relation for the kinetic energy and,
furthermore, it generates contributions to the nucleon effective mass
stemming from vector meson exchange. On the other hand, using the Bonn
potential
parameters with a ``Breit reduced'' potential would not be, strictly speaking,
consistent. While this may not be a major problem in practice (at least in the
case of $m_N^*$ the dependence on the choice of parameters is rather mild), a
better option might turn out to be the one of fixing the parameters on some
nuclear properties.

The results presented here indicate that at the mean field level it is possible
to map, with high accuracy, the many-body relativistic field theoretical
problem to an equivalent one, expressed within the standard hamiltonian
formulation. The next step would be, of course, to verify whether such a
correspondence is valid also for correlations beyond the mean field, like,
e.~g., those described by the random phase approximation (RPA).
Calculations based on both non-relativistic\cite{Shi88} and
relativistic\cite{Bar95a} Hamiltonians have in fact shown the importance of the
exchange terms in the RPA series, whereas fully relativistic calculations have
always been performed keeping only the direct (ring) contributions.


\begin{references}

\bibitem{Ser86}       B. D. Serot and J. D. Walecka,
                      Adv.\ Nucl.\ Phys.\ {\bf 16}, 1 (1986).

\bibitem{Cel86}       L. S. Celenza and  C. M. Shakin,
                      {\em Relativistic Nuclear Physics}
                      (World Scientific, Singapore, 1986).

\bibitem{Ser92}       B. D. Serot,
                      Rep.\ Prog.\ Phys.\ {\bf 55}, 1855 (1992).

\bibitem{Bar95a}      M. B. Barbaro, A. De Pace, T. W. Donnelly,
                      and A. Molinari,
                      in press in Nucl.\ Phys.\ {\bf A} (1995).

\bibitem{For95}       J. L. Forest, V. R. Pandharipande, and J. L. Friar,
                      Phys.\ Rev.\ C {\bf 52}, 568 (1995);
                      J. L. Forest, V. R. Pandharipande, J. Carlson, and
                      R. Schiavilla,
                      Phys.\ Rev.\ C {\bf 52}, 576 (1995);

\bibitem{Lan71}       L. D. Landau and  E. M. Lifschitz,
                      {\em Course of Theoretical Physics}
                      (Pergamon, Oxford, 1971) Vol.~4.

\bibitem{Bro69}       G. E. Brown, A. D. Jackson, and T. S. Kuo,
                      Nucl.\ Phys.\ {\bf A133}, 481 (1969).

\bibitem{Erk74}       K. Erkelenz,
                      Phys.\ Rep.\ {\bf 13C}, 191 (1974).

\bibitem{Mac87}       R. Machleidt, K. Holinde, and Ch. Elster,
                      Phys.\ Rep.\ {\bf 149}, 1 (1987).

\bibitem{Fol50}       L. L. Foldy and S. A. Wouthuysen,
                      Phys.\ Rev.\ {\bf 78}, 29 (1950).

\bibitem{Bjo64}       J. D. Bjorken and  S. D. Drell,
                      {\em Relativistic Quantum Mechanics}
                      (McGraw-Hill, New York, 1964).

\bibitem{Itz80}       C. Itzykson and J.-B. Zuber,
                      {\em Quantum Field Theory}
                      (McGraw-Hill, New York, 1980).

\bibitem{Haj80}       Ch. Hajduk and P. U. Sauer,
                      Phys.\ Rev.\ C {\bf 22}, 3 (1980).

\bibitem{Hol81}       K. Holinde,
                      Phys.\ Rep.\ {\bf 68}, 121 (1981).

\bibitem{Mac86}       R. Machleidt,
                      in {\em Relativistic Dynamics and Quark-Nuclear Physics},
                      edited by M.~B. Johnson and A. Picklesimer
                      (Wiley, New York, 1986) p.~71.

\bibitem{Shi88}       T. Shigehara, K. Shimizu, and A. Arima,
                      Nucl.\ Phys.\ {\bf A477}, 583 (1988).

\end{references}
\end{document}